\documentclass[aps,prl,superscriptaddress,groupedaddress,twocolumn,showpacs,floatfix]{revtex4}
\usepackage{graphicx}
\usepackage{amsmath,amsfonts,amssymb,amsthm,verbatim}
\usepackage{pstricks} 
\usepackage[ansinew]{inputenc}
\usepackage{color}
\usepackage{epsfig}
\usepackage{natbib}

\newcommand{\be}{\begin{equation}}
\newcommand{\ee}{\end{equation}}
\newcommand{\bea}{\begin{eqnarray}}
\newcommand{\eea}{\end{eqnarray}}

\newcommand{\up}{\uparrow}
\newcommand{\down}{\downarrow}


\begin{document}

\title{Half-metal phases in a quantum wire with modulated spin-orbit interaction }

\author{D. C.~Cabra}
\affiliation{IFLP-CONICET, Departamento de F\'isica, Universidad
Nacional de La Plata,\\ CC 67 1900 La Plata, Argentina}
\author{G. L. Rossini}
\affiliation{IFLP-CONICET, Departamento de F\'isica, Universidad
Nacional de La Plata,\\ CC 67 1900 La Plata, Argentina}
\author{A.~Ferraz}
\affiliation{International Institute of Physics - UFRN, \\Department
of Experimental and Theoretical Physics - UFRN, Natal, Brazil}
\author{G. I.~Japaridze}
\affiliation{Ilia State University, Faculty of Natural Sciences and
Engineering, Cholokashvili avenue 3-5, 0162 Tbilisi
Georgia and\\
Andronikashvili Institute of Physics, Tamarashvili str. 6, 0177
Tbilisi, Georgia}
\author{H.~Johannesson}
\affiliation{Department of Physics, University of Gothenburg, SE 412
96 Gothenburg, Sweden}

\begin{abstract}

We propose a spin valve device based on the interplay of a modulated spin-orbit interaction and a uniform external magnetic field acting on a quantum wire. Half-metal phases, where electrons with only a selected spin polarization exhibit ballistic conductance, can be tuned by varying the magnetic field. These half-metal phases  are proven to be robust against electron-electron repulsive interactions. 
Our results arise from a combination of explicit band diagonalization, bosonization techniques and extensive DMRG computations.
\end{abstract}


\date{\today}

\maketitle


The ability to control and manipulate electron spins with an
efficiency comparable to that of present-day (charge) electronics is one of the 
major goals of modern spintronics
[\onlinecite{Wolf_et_al_Science_01,Fabian_Zutic_Review_04,Fabian_Zutic_Review_09}].
As it comes to applications, fabrication of a device providing spin-dependent
currents is a central issue.
[\onlinecite{Bercioux_Lucignano_2015}]. A seminal blueprint for a spin-filtering scheme
was proposed in a paper published more than two decades ago [\onlinecite{Datta_Das}].  
The Datta-Das transistor uses a conductor contacted to a ferromagnetic source and drain,
subject to a Rashba spin-orbit interaction (SOI) [\onlinecite{Rashba}]. 
Depending on the spin orientations in the source and in the drain, the current 
flowing through the device can be controlled by rotating the spins of the incoming electrons, using a
gate voltage to tune the strength of the SOI [\onlinecite{DD_Transistor_1,DD_Transistor_2}]. 
Progress notwithstanding [\onlinecite{Chuang}], reliable injection of spin-polarized electrons from a ferromagnet 
into a semiconductor remains a challenge, and so does the very realization of a functional device producing 
spin-dependent currents.  

\begin{figure}[ht]
\begin{centering}
\includegraphics[scale=0.50]{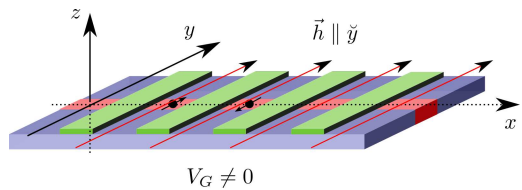}
\par\end{centering}
\caption{A qualitative sketch of the quantum
device discussed in this Letter. A quantum wire supporting Rashba SOI is gated
by a periodic sequence of equally charged top
gates. A transverse uniform magnetic field is applied in the direction
perpendicular to the current (wire) and the external electric field.} \label{fig:1}
\end{figure}

Because a SOI couples spin and momentum of
charge carriers, it also provides a differentiated effect on each spin
polarization. This opens a window for 
a spin-filtering regime, where only electrons with one spin polarization
carry current, while electrons with the opposite spin polarization are
gapped. This possibility is most profoundly displayed in the
case of one-dimensional conductors where a uniform Rashba SOI leads to a
spin-dependent shift of the electron dispersion by a momentum $\tau
q_{0}$, with $\tau = +,-$  the spin polarization along an axis
determined by the SOI [\onlinecite{streda:2003}]. 
A Peierls-type mechanism for a spin-based current switch was
identified in [\onlinecite{JJF_Paper_09}], where it was shown that
a {\em  spatially smooth} modulated Rashba SOI coupling opens
both charge and spin gaps in the system at commensurate band fillings. 
Such an interaction could be generated by a periodic gate
configuration schematically shown in Fig.~(\ref{fig:1}). 
%
In subsequent studies the effect of induced charge density wave
correlations in the quantum wire due to the periodic gating
was examined,
and the optimal regime where %
insulating current blockade occurs was
determined [\onlinecite{MGJJ_Paper_11}]. Other aspects of 
1D electron transport in the presence of modulated Rashba interactions have also been
discussed in the literature \cite{REVIEW}.

In this Letter $-$ building on the Peierls-type mechanism identified in [\onlinecite{JJF_Paper_09}] $-$
 we show how the interplay between a spatially smooth modulated Rashba SOI
and an applied magnetic field along the SOI axis may induce a 
selective opening of energy gaps, providing for spin-polarized electron conductance in a quantum wire. 
A detailed analysis, based on explicit band diagonalization, bosonization, and extensive DMRG computations, 
proves that the resulting half-metal phases are stable against repulsive electron-electron interactions, suggesting
that the proposed scheme can be realized in the laboratory.

%

To elucidate the physics underlying the magnetically controlled half-metal phases, we
set out by explaining how a uniform transverse magnetic field parallel to an SOI 
axis can generate a spin-selective metal-insulator transition. We then specialize
 to the case of commensurate magnetization and band filling, where
one spin projection electron subsystem is pinned by the interactions 
while the other remains gapless and as a result the system shows a perfect spin-filtering effect. 


Using a tight-binding formalism, with the spin-orbit coupled electrons confined to a single 1D conduction channel,
we model the system by the Hamiltonian  
\begin{eqnarray}\label{Hamiltonian_1}
H \!=\! &-&t \sum_{n,\alpha}\! \left(
c^{\dagger}_{n,\alpha}c^{\phantom{\dagger}}_{n+1,\alpha}
\!+\!\mbox{H.c.}\right)
\!-\! \mu\sum_{n, \alpha}\rho_{n,\alpha} \nonumber\\
&-&i\!\sum_{n,\alpha,\beta}\gamma_{R}(n)\!\left[c^{\dag}_{n,\alpha}
\sigma^{y}_{\alpha\beta} c^{\phantom{\dag}}_{n\!+\!1,\beta} +
\mbox{H.c.}\right]\nonumber\\
&+&\frac{h_{y}}{2}\sum_{n,\alpha,\beta}c^{\dag}_{n,\alpha}
\sigma^{y}_{\alpha\beta} c^{\phantom{\dag}}_{n,\beta},
\end{eqnarray}
for now ignoring the electron-electron interaction. Here $c^{\dagger}_{n,\alpha}$ ($c^{\phantom{\dag}}_{n,\alpha}$) is
the creation (annihilation) operator for an electron with spin
${\alpha}=\uparrow,\downarrow$ on site $n$,
$\rho_{n,\alpha}=c^{\dagger}_{n,\alpha}c^{\phantom{\dag}}_{n,\alpha}$,\,
$t$ is the electron hopping amplitude, $\mu$ a chemical potential,
$h_y$ is the  external magnetic field along the SOI axis $\sim \hat{y}$. 
The second line in (\ref{Hamiltonian_1}), with $\gamma_{R}(n)=\gamma_0+\gamma_1 \cos(Qn)$,  introduces the Rashba SOI 
with $\gamma_0 \ (\gamma_1)$ 
being the
amplitude of its uniform (modulated) part. For transparency and ease of notation we have omitted the modulation of
the chemical potential term caused by the modulated gate potential. It can be shown to result only in a band gap renormalization, 
an effect which can easily be included {\em a posteriori} by following a prescription in [\onlinecite{MGJJ_Paper_11}]. 

Choosing $\hat{y}$ as spin quantization axis, 
the uniform part of the SOI in (\ref{Hamiltonian_1}) is seen to split the dispersion relations of the rotated spins by a momentum $\tau q_{0}$, with $q_{0}=\arctan(\gamma_0/t)$ and with $\tau=\pm$ the spin projections along $\hat{y}$. In addition, the $\tau=\pm$ bands are split also by a Zeeman energy $\Delta \varepsilon=-\tau h_y/2$  due to the magnetic field $h_y$. 
For a given filling fraction $\nu$ and magnetization $m$, the right and left Fermi momenta for each band are given by
\begin{eqnarray}
&k_{F,\tau}^{R/L}  =  \tau q_{0} \pm k^{0}_{F,\tau}\,& 
\end{eqnarray}
where $k_{F,\tau}^{0}  =  (\nu+\tau m)\pi/2$.

To assess the impact of the modulated part of the SOI in (\ref{Hamiltonian_1}), it is convenient to use a bosonization approach [\onlinecite{Giamarchi_book}]. This will also be practical when we later analyze the role of electron-electron interactions, with bosonization offering an expedient view on correlation effects in the presence of a Rashba SOI  [\onlinecite{SOI}]. Introducing Bose fields $\varphi_{\tau}$ and their duals $\vartheta_{\tau}$, standard bosonization maps the low-energy sector of the Hamiltonian in (\ref{Hamiltonian_1}) to an effective continuum theory
$H^{bos}=H^{bos}_{0}+H^{bos}_{\gamma_{1}}$, where
%
\begin{widetext}
\begin{eqnarray}
H_{0}^{bos}&=&\frac{v_{F,\tau}}{2} \sum_{\tau=\pm} \int dx\
\left[\left(\partial_{x}\varphi_{\tau}\right)^{2}+
\left(\partial_{x}\vartheta_{\tau}\right)^{2}\right]\, ,\label{eq:H-0-bos}\\
H_{\gamma_{1}}^{bos}&=&-\frac{M_R}{\pi a_{0}}\sum_{\tau=\pm} \sum_{j=\pm 1}
\!\int dx 
\sin\left[(2k^{0}_{F,\tau}+jQ)x+k^{0}_{F,\tau}
+\sqrt{4\pi}\varphi_{\tau}(x)\right]\, ,
\label{eq:H-gamma1-bos}
\end{eqnarray}
\end{widetext}
Here $v_{F,\tau}=2\sqrt{t^2+\gamma_0^ 2}\sin(k^{0}_{F,\tau})$ are the Fermi velocities, $M_R = \gamma_1 \sin (q_0 a_0)$
measures the strength of the modulated Rashba SOI, and $a_0$ is a short-distance cutoff conveniently taken as the lattice spacing.
Note that the bosonized Hamiltonian can also be decomposed as $H^{bos} = H^{bos}_{+}+H^{bos}_{-}$, 
with each piece containing only Bose fields with either $\tau=+$ or $\tau=-$, showing that the rotated spins are good quantum numbers.

From Eq.~(\ref{eq:H-gamma1-bos}) one concludes that 
the modulated Rashba SOI can have an effect only at commensurate band fillings 
given by the conditions
\begin{equation}
 2k^{0}_{F,\tau} \pm Q \approx 0 \,\,\text{ mod }2\pi\, .
\label{eq:condition.1}
\end{equation}
since for all other cases the sine-term in (\ref{eq:H-gamma1-bos}) oscillates rapidly and vanishes upon integration.
At finite magnetization the commensurability conditions in (\ref{eq:condition.1})
are \emph{different} for each spin projection;
when the conditions are met for a given polarization, a relevant perturbation (in the sense of the renormalization group [\onlinecite{Giamarchi_book}])
is present, opening a gap to the corresponding electron excitations. 
To be precise, when only the spin ``+'' sector satisfies the commensurability condition,
$H^{bos}_{+}$ is a massive sine-Gordon model and becomes gapful 
with a mass gap $M_R$,
while $H^{bos}_{-}$ describes a gapless Gaussian model.
Then the ``+'' spin electron subsystem is pinned by the
commensurability effect in a long-range ordered quantum configuration while the ``$-$'' sector remains gapless and disordered.
Charge transport in the gapped sector is forbidden while in the gapless one transport is ballistic.
Therefore the wire displays a {\em half-metal} behavior [\onlinecite{Groot}] and acts as a spin filter.
Such a half-metal phase,  induced by a magnetic field and a modulated gate voltage, might be experimentally realized and controlled 
by varying the electron chemical potential via a backgate.
As the conducting sector could be turned ON/OFF or even changed to the other spin polarization, 
the proposed device would be properly called a magnetic spin valve.

%
\begin{figure}[ht]
\begin{centering}
\includegraphics[scale=0.35]{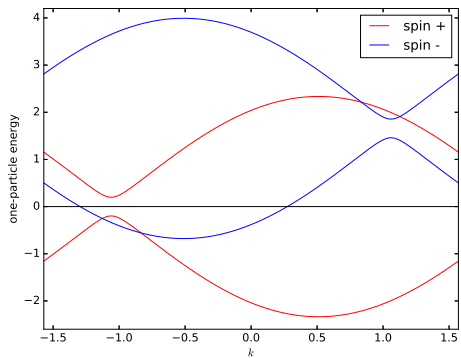}
\par\end{centering}
\caption{Single-particle dispersion relation for $Q=\pi$, $\nu=3/4$,
$m=1/4$, with Fermi level $\epsilon_F=0$. The lower band with spin ``+'' is completely filled, but the
lower band with spin ``-''  is partially filled. There is a gap to
charge excitations with spin ``+'', but no gap to charge excitations
with spin ``-''. Rashba coefficients are taken as $\gamma_0/t=\tan(\pi/6)$ and $\gamma_1/t=0.2$}
\label{fig:2}
\end{figure}
The results above, obtained for non-interacting electrons, can easily be checked numerically by explicit band diagonalization.
In Fig.~\ref{fig:2} we illustrate a simple case by plotting the
single-particle dispersion relations for Rashba modulation $Q=\pi$, 
with band filling $\nu=3/4$ and magnetization $m=1/4$. 

To find out whether the half-metal phases identified above can be realized experimentally, it is crucial to analyze the effect of electron-electron correlations.  
For this purpose, we here model the screened Coulomb interaction between electrons by an on-site Hubbard interaction
\begin{equation} \label{Hubbard}
H_U = U \sum_{n}\rho_{n,\up}\rho_{n,\down},
\end{equation}
to be added to the Hamiltonian in (\ref{Hamiltonian_1}).
Its bosonized expression in the rotated basis (with spin projections $\pm$ along the $\hat{y}$-axis) takes the form
\begin{widetext}
\begin{equation} 
H^{bos}_{U}=\frac{U}{\pi}\int dx\big[
\partial_{x}\varphi_{+}\partial_{x}\varphi_{-}+
\frac{1}{\pi a^{2}_{0}}\sin(\sqrt{4\pi}\varphi_{+}+2k^{0}_{F,+}x)
\sin(\sqrt{4\pi}\varphi_{-}+2k^{0}_{F,-}x)\}. \label{eq:H-U-bos}  
\end{equation}
\end{widetext}
%

%
%
In a half-metal phase, where condition (\ref{eq:condition.1}) is satisfied in one spin sector, 
the sine factors in ({\ref{eq:H-U-bos}) come out incommensurate, implying that their product averages to
zero upon integration. This leaves us with the gradient term in ({\ref{eq:H-U-bos}), which, however, 
is exactly marginal and therefore cannot close the gap $\sim M_R$.
One concludes that electron correlations 
do not destabilize the magnetically controlled half-metal phases at the low energies where bosonization applies.

What about intermediate energies where effects from the lattice may play a role? To find out, we have carried 
out large-scale DMRG computations on the Hamiltonian in (\ref{Hamiltonian_1}) with the Hubbard interaction (\ref{Hubbard}) added, $H'=H+H_U.$ 
The computations were performed in the same rotated spin basis as used above, with electron operators
\begin{equation} \label{spinor}
 \left( \begin{array}{c}
 d_{n,+} \\
d_{n,-} \end{array} \right) \equiv \frac{1}{\sqrt{2}} \left(
\begin{array}{c}
c_{n,\uparrow} - ic_{n,\downarrow} \\
-ic_{n,\uparrow} + c_{n,\downarrow}
\end{array} \right).
\end{equation}
The Hamiltonian $H'$ commutes with the total
charge operator $\sum_{n,\tau}
d^{\dagger}_{n,\tau}d^{\phantom{\dagger}}_{n,\tau} $ and the total
spin $y$-component operator $\frac{1}{2} \sum_{n,\tau}
\tau d^{\dagger}_{n,\tau}d^{\phantom{\dagger}}_{n,\tau}$. As a
consequence, the eigenvalues of ${\hat N}_{+}=\sum_{n}
d^{\dagger}_{n,+}d^{\phantom{\dagger}}_{n,+}$ and ${\hat N}_{-}
=\sum_{n} d^{\dagger}_{n,-}d^{\phantom{\dagger}}_{n,-}$ are good
quantum numbers describing the occupation of states with each spin
projection $\tau=\pm$.
For a  chain of length $L$ with band filling $\nu$ and magnetization $m$ 
we then consider the lowest energy state
in the subspace with
$N_{+}=L(\nu+m)/2$  occupied states with  spin ``+''  and
$N_{-}=L(\nu-m)/2$  occupied states with  spin  ``-'', denoting by
$E_{0}(N_{+},N_{-})$ the corresponding ground state energy.
One-particle excitation gaps $\Delta_{\pm}$
are defined as the average energy cost of adding or removing an electron with a given spin projection $\pm$ [\onlinecite{Manmana}],
\begin{equation}
2\Delta_{\pm}\!=\!E_{0}(N_{\pm}\!+\!1,N_{\mp})\!+\!E_{0}(N_{\pm}\!-\!1,N_{\mp})\!-\!2E_{0}(N_{+},N_{-}), \nonumber
\label{eq:delta+}
\end{equation}
and coincide, in the gapped spin sector of a half-metal phase, with the excitation gap $M_R$ of the massive sine-Gordon model above.
Importantly, this is the gap that determines the current blockade effect in our proposed spin valve device. 
%

In Fig.~(\ref{fig:magnetized-1p}) we show our DMRG results for the one-particle gaps and 
their infinite length extrapolation in the half-metal phase sustained by a
Rashba SOI modulation $Q=\pi$, filling $\nu=3/4$ and magnetization $m=1/4$,
with the condition (\ref{eq:condition.1}) satisfied for spin ``+'', and
with the repulsive Hubbard interaction
ranging from $U=0$ to $U=25\, t$. The remaining Hamiltonian parameters 
were set to $\gamma_0/t=\tan(\pi/6)$ and $\gamma_1/t=0.2$. The computations were carried 
out for finite-length systems with $L=48$, $64$ and $96$ sites, using the ALPS library [\onlinecite{Bauer:2011}]. 
Most of the
data points have been obtained keeping $800$ states during $30$
sweeps. The estimated error for energy measures is $10^{-3}\,t$, which
ensures enough precision for the gaps we report.

In the non-interacting case, as discussed above,
the spin ``+'' band is half-filled and gapped (with $\Delta_{+} =0.2 t$ in Fig.~(\ref{fig:magnetized-1p})) while the spin ``-'' band
is quarter-filled and gapless. As seen in Fig.~(\ref{fig:magnetized-1p}), electron-electron repulsion $\sim U$ 
reduces the gap $\Delta_{+}$, however, without closing it for any $U$.
On the contrary, the system seems to stabilize with a different gap in the large $U$ limit. 
The spin ``-'' gap, which vanishes at $U=0$, also scales to zero
for any $U$. This last result, however, is highly sensitive to finite-size
effects; in particular, the dispersion seen in Fig.~(\ref{fig:magnetized-1p})} at $U=0$ is due to
the incommensurability between the band energy minimum of the shifted bands
and the discrete finite-length reciprocal lattice.
\begin{figure}
\begin{centering}
\includegraphics[width=8cm]{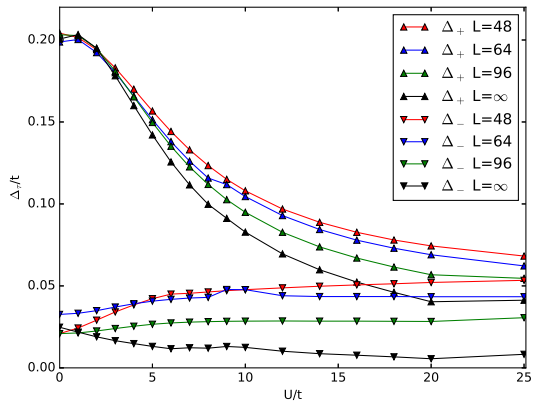}
\par\end{centering}
\caption{DMRG results for one-particle gaps as function of $U$ in the half-metal phase depicted in Fig.~\ref{fig:2}.  }
\label{fig:magnetized-1p}
\end{figure}

It is instructive to consider also the two-particle excitation gaps which describe
pure charge or  pure spin excitations, related to the bosonic charge and spin fields
$\varphi_c = (\varphi_{+} + \varphi_{-})/\sqrt{2}$ and $\varphi_s = (\varphi_{+} - \varphi_{-})/\sqrt{2}$
respectively}.
The charge gap $\Delta_c$ is
defined by 
%
\begin{multline}
\Delta_{c}=\frac{1}{2}[E_{0}(N_{+}\!+\!1,N_{-}\!+\!1) \\ +E_{0}(N_{+}\!-\!1,N_{-}\!-\!1)-2E_{0}(N_{+},N_{-})],
\label{eq:deltaC}
\end{multline}
while the spin gap $\Delta_s$ is defined by 
%
\begin{multline}
\Delta_{s}=\frac{1}{2}[E_{0}(N_{+}\!+\!1,N_{-}\!-\!1) \\ +E_{0}(N_{+}\!-\!1,N_{-}\!+\!1)-2E_{0}(N_{+},N_{-})].
\label{eq:deltaS}
\end{multline}
%
For the non-interacting system ($U=0$), the two-particle
gaps are simply related to the gaps of single particles as 
$\Delta_{c}=\Delta_{s}=\Delta_{+}+\Delta_{-}$. The presence of
electron interactions may change these relations, however. In fact, the more different
the charge and spin gaps are, the more correlated the system is, making two-particle
gaps sensitive probes of correlation effects.

\begin{figure}
\begin{centering}
\includegraphics[width=8cm]{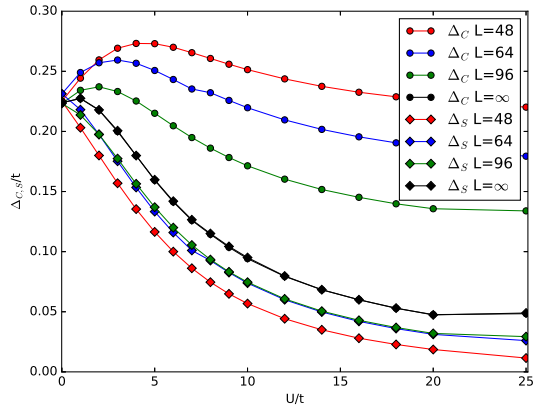}
\par\end{centering}
\caption{DMRG results for two-particle gaps as function of $U$ in the half-metal phase depicted in Fig.~\ref{fig:2}.}
\label{fig:magnetized-2p}
\end{figure}

In Fig.~(\ref{fig:magnetized-2p}) we present the corresponding numerical results for the two-particle gaps.
Although the DMRG data show a marked size dependence, 
the infinite-length extrapolation following a $1/L$ law fits remarkably well the finite-size data,
showing that the charge and spin gaps  remain coincident at any $U$, 
being the sum of the one-particle gaps. 
This strongly supports the picture of the system remaining in the same 
non-correlated phase as when $U=0$. For weak and intermediate electron-electron
interactions, with $U \lesssim t$, the only noticeable interaction effect is a small reduction
of the single-particle gap (cf. Fig.~(\ref{fig:magnetized-1p})).

Having furnished a proof-of-concept for a novel type of spin valve device $-$ exploiting the
possibility of magnetically controlled half-metal phases in a quantum wire subject 
to periodic gating $-$ what are the prospects to actually make it work? 
While an exhaustive analysis goes beyond the scope of this Letter, let us attempt a
brief appraisal. Given that correlation effects are negligible for the weak to intermediate
interaction strength $U/t \lesssim {\cal O}(1)$ expected for a gated quantum wire supported by a semiconductor
heterostructure [\onlinecite{Fulde}], the key parameter that determines the functionality of the device 
is the single-particle gap $M_R$, defined above for noninteracting electrons as $M_R=2\gamma_1\sin(q_0 a_0)$. 
When including the effect from the 
modulation of the chemical potential due to the periodic gating (cf. Fig. (\ref{fig:1})), $M_R$ gets ``dressed" by the amplitude $\mu_{\text{mod}}$ of the modulation
and is replaced by  
\begin{equation} \label{dressed}
M_{R,\mu_{\text{mod}}} = \sqrt{M_R^2 \pm \mu_{\text{mod}}M_R\cos(\pi \nu) + \mu^2_{\text{mod}}/4}, 
\end{equation}
with $\nu$ the band filling, and with the sign $+ \ (-)$ coding for the Rashba and chemical potential modulations
being out-of-phase (in-phase) depending on material and design of the setup [\onlinecite{MGJJ_Paper_11}].
As a case study, we use data obtained from an experiment on gate-controlled Rashba interaction in a square asymmetric
InAs quantum well [\onlinecite{Grundler}], assuming that it has been gated to define a single-channel micron-range ballistic 
quantum wire. Combined with data from [\onlinecite{Bhattacharaya}], we obtain that 
$\gamma_1 \approx 1\times 10^{-11}$ eVm, $q_0 a_0 \approx 0.1$, 4 meV $\le \mu_{\text{mod}} \le$ 10 meV, and $\nu \approx
0.04$. With the chemical potential modulation here being out of phase with that of the Rashba SOI [\onlinecite{MGJJ_Paper_11}], 
Eq. (\ref{dressed}) yields the estimate
\begin{equation}
0.3 \ \mbox{meV} \le \ M_{R,\mu_{\text{mod}}} \le 3.0 \ \mbox{meV}.
\end{equation}
To prevent thermal leakage across the single-particle gap that
serves to blockade transport of electrons with ``wrong" spin, a
device based on the same materials and basic architecture as in [\onlinecite{Grundler}] would thus have to operate well below 1K. 
Functionality of a device at higher temperatures may be achieved by boosting the effective Rashba couplings [\onlinecite{NatMat}], or, maybe more workable,
by "band engineering", using composite materials where the Rashba and chemical potential modulations are in-phase instead of out-of-phase.  

In summary, in this Letter we have shown that a combination of
a uniform magnetic field and a gate voltage controlled modulated Rashba SOI 
may drive a quantum wire into half-metal phases, with transport only of electrons 
with a given spin polarization. We have identified the commensurability 
conditions for the appearance of such phases, and also provided analytical and
numerical evidence for their robustness against electron-electron interactions.
Our results hold promise for the design of a
magnetic field-controlled spin valve device, without resorting to injection from 
ferromagnetic leads. To assess the viability and functionality of such a design requires further work, 
theoretical as well as experimental. \\

\begin{acknowledgments} 
This work was partially supported by CONICET (PIP 2015-813) and ANPCyT (PICT
2012-1724), Argentina, the Brazilian Ministry of Education and the CNPq, the Shota Rustaveli Georgian National Science
Foundation through the grant FR/265/6-100/14, and by the Swedish Research Council 
through grant no. 621-2014-5972.
\end{acknowledgments}

\bibliographystyle{plain}


\end{document}